\documentclass[aps,pra,preprint,superscriptaddress]{revtex4-1}

\usepackage{amsmath,amsfonts,amssymb}
\usepackage{xcolor,graphicx}
\usepackage{braket}
\usepackage{float}
\usepackage[pdftex]{hyperref}
\usepackage{csquotes}
\hypersetup{
	colorlinks = true,
	linkcolor = blue,
	anchorcolor = blue,
	citecolor = red,
	filecolor = blue,
	urlcolor = blue}

\begin{document}

\title{Some non-algebraic forms of $\exp(A+B)$}

\author{Marco A. Tapia-Valerdi}
\email[e-mail: ]{mvalerdi@inaoep.mx}
\affiliation{Instituto Nacional de Astrofísica Óptica y Electrónica (INAOE)\\ Luis Enrique Erro 1, Santa María Tonantzintla, Puebla, 72840, Mexico}
\author{I. Ramos-Prieto}
\email[e-mail: ]{iran@inaoep.mx}
\affiliation{Instituto Nacional de Astrofísica Óptica y Electrónica (INAOE)\\ Luis Enrique Erro 1, Santa María Tonantzintla, Puebla, 72840, Mexico}
\author{F. Soto-Eguibar}
\email[e-mail: ]{feguibar@inaoep.mx}
\affiliation{Instituto Nacional de Astrofísica Óptica y Electrónica (INAOE)\\ Luis Enrique Erro 1, Santa María Tonantzintla, Puebla, 72840, Mexico}
\author{H. M. Moya-Cessa}
\email[e-mail: ]{hmmc@inaoep.mx}
\affiliation{Instituto Nacional de Astrofísica Óptica y Electrónica (INAOE)\\ Luis Enrique Erro 1, Santa María Tonantzintla, Puebla, 72840, Mexico}

\begin{abstract}
We present examples where expressions for $\exp(\hat{A}+\hat{B})$ can be derived even though the operators (or superoperators) $\hat{A}$ and $\hat{B}$ do not commute in a manner that leads to known factorizations. We apply our factorization to the case of a Lindblad operator modeling single photon decay and to a binary Glauber-Fock photonic lattice.
\end{abstract}

\date{\today}

\maketitle

\section{Introduction}
In various fields of physics, such as classical mechanics \cite{yoshida1993recent,mclachlan2002splitting}, electrodynamics \cite{plass1961classical}, and quantum mechanics \cite{tapia2024generalization}, it is common to encounter equations of the form $\frac{\partial x}{\partial t}= (\hat{A}+\hat{B})x$. In quantum mechanics, this can be readily identified with the Schrödinger equation, whose formal solution involves an evolution operator that can be separated into holonomy and dynamic operators \cite{yu2023evolution}. The solutions to many physical evolution equations often take the form of an exponential function involving two or more operators acting on a given initial condition; thus, it is frequently necessary to find expressions for $\exp(\hat{A}+\hat{B})$. When operators $\hat{A}$ and $\hat{B}$ commute, the adoption of Lie algebras has become widespread among researchers investigating nonclassical behaviors of light in quantum-optical systems \cite{teuber2020solving}. This approach facilitates the creation of a product of exponentials \cite{sack1958taylor,wei1963lie,wilcox1967exponential}, and even decomposition formulas of exponential operators in Banach and Lie algebras \cite{suzuki1985decomposition,ban1993decomposition}, with applications in quantum Monte Carlo techniques and structured light. However, when $\hat{A}$ and $\hat{B}$ are non-commuting operators, this task can be challenging or even impossible, as in cases where $[\hat{A},\hat{B}]=\hat{A}\hat{B}$ or $[\hat{A},\hat{B}]=2\hat{A}\hat{B}$. The Baker-Campbell-Hausdorff formula \cite{louisell} is not applicable due to additional commutation relations that arise, preventing a closed-form solution using established operator algebras~\cite{RossmannW,Hall_2013}. We begin by examining cases where the operators satisfy two commutation rules, deriving expressions for the evolution operators and applying these to specific physical scenarios. These specific scenarios include a master equation in its Lindblad form and the Glauber-Fock photonic lattice. In the following two paragraphs, we introduce these two topics to provide a brief, but complete outline.

A significant challenge in quantum mechanics is describing systems that interact with their environments, leading to dissipative or absorptive interactions. When a quantum system interacts with its environment at arbitrary temperatures, master equations emerge as fundamental tools for this description. Master equations formulated in Lindblad form provide the most comprehensive description of open quantum systems under the assumption of Markovian dynamics \cite{breuer2002theory,carmichael1993open,manzano2020short}. The Lindblad master equation is composed of a Hermitian Hamiltonian segment, governing the coherent evolution of the system, and the Lindblad dissipators, which capture the system's coupling with the environment, describing the progressive loss of energy, coherence, and information. Traditional methods of handling these equations include transforming them into Fokker–Planck equations \cite{risken1984solutions} or using Langevin equations \cite{gardiner1985handbook}. However, these methods often present difficulties when applying solutions to arbitrary initial conditions. In contrast, superoperator techniques offer direct application to an initial wave function, though they are part of the non-Hermitian Hamiltonian formalism \cite{ashida2020non}. The non-Hermitian nature of open systems, once seen as a hindrance, is now recognized for its potential to reveal fascinating effects. This has spurred interest in research areas such as parity-time $\mathcal{PT}$ symmetry, quasi-$\mathcal{PT}$ systems \cite{long2022non,bender1998real,hernandez2023exact,longhi2020quantum}, electrical circuits, and exceptional points \cite{liu2021non,liu2023experimental,minganti2019quantum,wiersig2020review}.

The discrete coupling or tunneling process between periodically arranged potential wells is another fundamental topic extensively studied across various branches of physics. In optics, arrays of weakly coupled waveguides serve as prime examples where the discrete diffraction properties of these configurations can predictably mold light flow, making coupling dynamics observable and investigable \cite{perez2010glauber}. Consequently, light propagation in waveguide lattices has garnered significant interest. These arrays provide a versatile platform for observing numerous processes \cite{vicencio2015observation,trompeter2006visual,perez2016generalized,Ancheyta_2017,Ramos_2021,Urzua_2024}, and exploring new applications. Lattice models, often used in condensed-matter physics, are powerful tools for studying crystalline structures \cite{altland2010condensed,kartashov2011solitons}, and have direct connections to computational physics \cite{schafer2020tools,lewenstein2012ultracold}.

The article is organized as follows: In Section \ref{AB}, we examine the case where two operators satisfy the commutation rule $[\hat{A},\hat{B}]=\hat{A}\hat{B}$. This allows us to derive an expression for $(\hat{A}+\hat{B})^k$, which is crucial for obtaining an evolution operator that can be easily applied to an arbitrary initial condition. We also provide an example involving a master equation in its Lindblad form, demonstrating how this technique can be used to derive the evolution of the density matrix. In subsections \ref{coherent state} and \ref{thermal states}, we apply these results to find an expression for the coherent state as the initial condition and determine the average number of photons for various initial conditions, such as thermal states and coherent states. In Section \ref{2AB}, we present another case where we calculate the exponential of the sum of two operators using their Taylor series representation. Given that these operators satisfy the commutation relation $[\hat{A},\hat{B}]=2\hat{A}\hat{B}$, we can express the result in terms of sines and cosines. We provide a specific example where the system's Hamiltonian operators follow the aforementioned commutation rule, allowing us to derive an analytical expression for its wave function, which can be associated with a Glauber-Fock photonic lattice. Our conclusions are presented in Section \ref{conclusiones}.

\section{Case $[\hat{A},\hat{B}]=\hat{A}\hat{B}$} \label{AB}
Assuming that operators $\hat{A}$ and $\hat{B}$ fulfill the commutation rule $[\hat{A},\hat{B}]= \hat{A}\hat{B}$, that implies $\hat{B}\hat{A}=0$, it is possible to use mathematical induction to prove that
\begin{equation}\label{201}
(\hat{A}+\hat{B})^k= \hat{B}^k + \sum_{m=1}^k \hat{A}^{m} \hat{B}^{k-m},   
\end{equation}
with $k$ any non-negative integer. \\

We now consider the Lindblad master equation \cite{carmichael1993open,breuer2002theory,manzano2020short} 
\begin{equation}
\frac{d\hat{\rho}(t)}{dt}=2\gamma\hat{A}\hat{\rho}\hat{A}^\dagger-\gamma\hat{\rho}\hat{A}^\dagger\hat{A}-\gamma\hat{A}^\dagger\hat{A}\hat{\rho},
\end{equation}
where $\hat{A}$ is an arbitrary operator. When $\hat{A}=\hat{a}$, i.e., the usual annihilation operator for the harmonic oscillator, we obtain the well known master equation that describes the decay of a quantized electromagnetic field under the Born-Markov approximation. Generalizations of  annihilation operators also may be considered, namely, nonlinear annihilation operators \cite{man1997f,de2015markovian,de2011f,Ramos_2014}, where the harmonic oscillator annihilation operator, $\hat{a}$, is replaced by a nonlinear one: $f(\hat{a}^{\dagger}\hat{a})\hat{a}$, being $f$ an arbitrary well behaved function. We replace $\hat{a}$ by $\hat{V}=\frac{1}{\sqrt{1+\hat{a}^{\dagger}\hat{a}}}\hat{a}$ and write the Lindblad master equation
\begin{equation}\label{LindbladME}
\frac{d\hat{\rho}(t)}{dt}=2\gamma\hat{V}\hat{\rho}\hat{V}^\dagger-\gamma\hat{\rho}\hat{V}^\dagger\hat{V}-\gamma\hat{V}^\dagger\hat{V}\hat{\rho},
\end{equation}
for the well-known London-Susskind–Glogower phase operators \cite{london1926jacobischen,susskind1964quantum} with $\gamma$ the decay constant; using the fact that $\hat{V}\hat{V}^\dagger=1$ and $\hat{V}^\dagger\hat{V}=1-\ket0\bra0$, we can rewrite the above expression as 
\begin{equation}\label{masterlondon}
\frac{d\hat{\rho}(t)}{dt}=2\gamma\hat{V}\hat{\rho}\hat{V}^\dagger+\gamma\hat{\rho}\ket{0}\bra{0}+\gamma\ket{0}\bra{0}\hat{\rho}-2\gamma\hat{\rho}.
\end{equation}
To solve Eq.~\eqref{masterlondon}, we define the superoperators $\hat{L}\hat{\rho}=\gamma\hat{\rho}\ket{0}\bra{0}+\gamma\ket{0}\bra{0}\hat{\rho}$ and $\hat{J}\hat{\rho}=2\gamma\hat{V}\hat{\rho}\hat{V}^\dagger$, in such a way that we can write
\begin{equation}
 \frac{d\hat{\rho}(t)}{dt} = (\hat{J}+\hat{L}-2\gamma)\hat{\rho}, 
\end{equation}
and we have a differential equation that we can easily solve formally by direct integration as
\begin{equation}\label{solution1}
\hat{\rho}(t)=e^{-2\gamma t}\exp\bigg[(\hat{J}+\hat{L})t\bigg]\hat{\rho}(0),
\end{equation}
where $\hat{\rho}(0)$ represents the density matrix at $t=0$. If we make the identifications $\hat{A} \rightarrow \hat{L}$ and $\hat{B}\rightarrow \hat{J}$, we observe that this two operators comply with the aforementioned commutation rule; thus, expanding the exponential into its Taylor series and using Eq.~\eqref{201}, allows us to write
\begin{equation}
(\hat{J}+\hat{L})^k\hat{\rho}(0)=\hat{J}^k\hat{\rho}(0)+\sum_{m=1}^k\hat{L}^m\hat{J}^{k-m}\hat{\rho}(0).
\end{equation}
In order to simplify the above equation, it is not difficult to prove by mathematical induction that
\begin{equation}
\begin{split}
\hat{J}^k\hat{\rho}(0)&=(2\gamma)^k\hat{V}^k\hat{\rho}(0)(\hat{V}^\dagger)^k,\\ \hat{L}^m\hat{\rho}(0)&=\gamma^m\bigg[\ket{0}\bra{0}\hat{\rho}(0)+(2^m-2)\ket0\bra{0}\hat{\rho}(0)\ket0\bra0+\hat{\rho}(0)\ket0\bra0\bigg],
\end{split}
\end{equation}
for any $k$ and $m$, non-negative integers, our solution \eqref{solution1} can then be written as
\begin{align}\label{solutionf}
e^{2\gamma t}\hat{\rho}(t)&=e^{\hat{J}t}\hat{\rho}(0)+\ket{0}\bra{0}\sum_{k=m}^\infty\frac{\gamma^k t^k}{k!}\sum_{m=1}^k2^{k-m}(2^m-2)\hat{V}^{(k-m)}\hat{\rho}(0)\hat{V}^{\dagger(k-m)}\ket{0}\bra{0}
\nonumber\\
&+\ket{0}\bra{0}\sum_{k=m}^\infty\frac{\gamma^k t^k}{k!}\sum_{m=1}^k2^{(k-m)}\hat{V}^{(k-m)}\hat{\rho}(0)\hat{V}^{\dagger(k-m)}\nonumber\\
&+\sum_{k=m}^\infty\frac{\gamma^k t^k}{k!}\sum_{m=1}^k2^{(k-m)}\hat{V}^{(k-m)}\hat{\rho}(0)\hat{V}^{\dagger(k-m)}\ket{0}\bra{0}.
\end{align}
As will be shown later, this equation can be employed to ascertain the temporal evolution of an arbitrary initial density matrix as well as to determine the average number of photons for certain initial conditions, such as coherent states and thermal states. The average number of photons is given by $\bar{n}(t)=\mbox{Tr}\bigg[\hat{a}^\dagger\hat{a}\hat{\rho}(t)\bigg]$; then, substituting our solution into this expression, using the fact that $\hat{n}\ket{0}\bra{0}=0$, and the cyclic property of the trace, we get that the average number of photons can be written as
\begin{equation}\label{numeropromedio}
\bar{n}(t)=e^{-2\gamma t}\mbox{Tr}\bigg[\hat{a}^\dagger\hat{a}e^{\hat{J}t}\hat{\rho}(0)\bigg].
\end{equation}
Next, we will consider as initial conditions a coherent state and a thermal state.

\subsection{Coherent states} \label{coherent state}
In appendix \ref{appendix a}, we prove that if we consider as initial condition a coherent state $\rho(0)=|\alpha\rangle\langle \alpha|$ in Eq. \eqref{solutionf}, we arrive to 
\begin{align}\label{solcoherente}
e^{2\gamma t}\hat{\rho}(t) &= e^{-|\alpha|^2}\sum_{n=0}^{\infty}\frac{(2|\alpha|\gamma t)^n}{n!}\sum_{j,j'=0}^{\infty}\frac{\alpha^j\alpha^{j'}}{\sqrt{(j+n)!(j'+n)!}}\ket{j'}\bra{j}\nonumber\\&+ e^{-|\alpha|^2}\ket{0}\bra{0}\sum_{m=1}^{\infty}\sum_{n=0}^\infty\frac{(\gamma t)^{n+m}}{(n+m)!}2^{n}(2^m-2)\frac{|\alpha|^{2n}}{n!}
\nonumber \\
& +e^{-|\alpha|^2}\sum_{m=1}^{\infty}\sum_{n=0}^\infty\frac{(\gamma t)^{n+m}}{(n+m)!}2^{n}\frac{|\alpha|^{2n}}{\sqrt{n!}}\sum_{j=0}^{\infty}\frac{\alpha^{*j}}{\sqrt{(j+n)!}}\ket{0}\bra{j}
\nonumber \\
&+e^{-|\alpha|^2}\sum_{m=1}^{\infty}\sum_{n=0}^\infty\frac{(\gamma t)^{n+m}}{(n+m)!}2^{n}\frac{|\alpha|^{2n}}{\sqrt{n!}}\sum_{j=0}^{\infty}\frac{\alpha^{j}}{\sqrt{(j+n)!}}\ket{j}\bra{0}.
\end{align}
An arbitrary initial density matrix may be  written as a superposition of coherent states with the help of the Glauber-Sudarshan $P$-function, i.e.,
\begin{equation}
    \hat{\rho}(0)=\frac{1}{\pi}\int d^2\alpha P(\alpha)\ket{\alpha}\bra{\alpha};
\end{equation}
therefore, by using the above expressions for coherent states, we can give a closed expression for arbitrary initial states.

\begin{figure}
\includegraphics[width = \linewidth ]{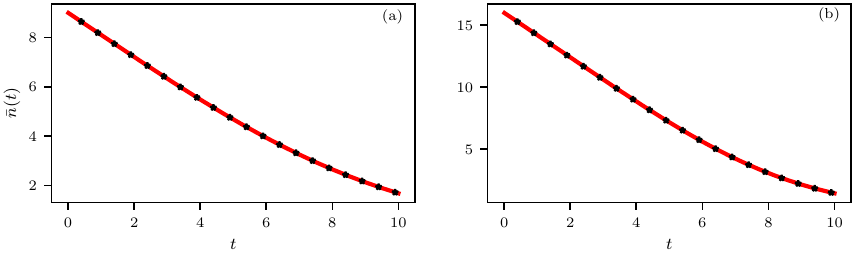}
\caption{The Lindblad master equation, Eq.~\eqref{LindbladME}, is solved numerically and compared with our analytical results for the expected photon number, given by Eq.~\eqref{npromedio 2}, which determines the expected photon number for a coherent state as the initial condition. The solid red curve represents the analytical solution, while the dots indicate the numerical results. Two sets of parameters are considered: (a) $\alpha=3$ and $\gamma=0.45$, (b) $\alpha=4$ and $\gamma=0.9$.
}
\label{coherentes}
\end{figure}

According to Eq.~\eqref{numeropromedio}, the average photon number is in this case
\begin{equation}
\bar{n}(t)=e^{-2\gamma t}\mbox{Tr}\Bigg[\sum_{n=0}^\infty\frac{(2\gamma t)^n}{n!}\hat{a}^\dagger\hat{a}\hat{V}^n\ket{\alpha}\bra{\alpha}\hat{V}^{\dagger n}\Bigg];
\end{equation}
furthermore, from Eq.~\eqref{acoherent}, we can obtain easily that
\begin{equation}
\bar{n}(t)=e^{-2\gamma t}e^{-|\alpha|^2}\sum_{n,k=0}^\infty\frac{(2\gamma t)^n|\alpha|^{2(n+k)}}{n!(n+k)!}k;
\end{equation}
we can multiply this last expression by the identity, written as $\bigg(\frac{\sqrt{8\gamma t|\alpha|^2}}{2}\bigg
)^{-k}\bigg(\frac{\sqrt{8\gamma t|\alpha|^2}}{2}\bigg)^{k}$, and introduce the well-known modified Bessel functions of the first kind~\cite{abramowitz1968handbook,arfken2011mathematical}
\begin{equation}
I_k(x)=\sum_{n=0}^\infty\frac{(x/2)^{k+2n}}{n!\Gamma(k+n+1)},
\end{equation}
being $\Gamma$ the gamma function. It is easy to see that
\begin{equation}\label{npromedio 2}
\bar{n}(t)=e^{-2\gamma t}e^{-|\alpha|^2}\sum_{k=0}^\infty k\left(\frac{|\alpha|^2}{2\gamma t}\right)^{k/2}I_{k}\left(\sqrt{8\gamma t|\alpha|^2}\right).
\end{equation}
In Fig.~\ref{coherentes}, we present the explicit evolution of the average photon number. Initially, the system is prepared in a coherent state. We analyze two distinct scenarios, each illustrating the decay of the initial average photon number over time. These scenarios vary in terms of the initial photon number $\bar{n}_0$ and the decay rate $\gamma$. As the decay rate $\gamma$ increases, the rate of decrease in the average photon number accelerates, driven by the exponential decay process even though the modified Bessel functions of the first kind have bottom-up behavior. Additionally, for longer time duration, we observe complete dissipation of the interaction into the environment, consistent with expectations.

\subsection{Thermal states}\label{thermal states}
\begin{figure}[H]
\centering
\includegraphics[width = \linewidth ]{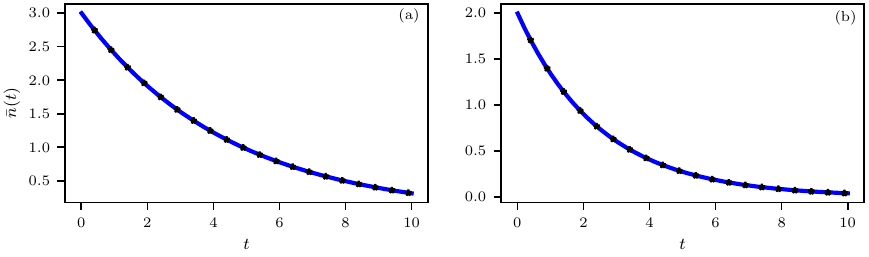}
\caption{We solve the Lindblad master equation, Eq.\eqref{LindbladME}, numerically and compare the results with our analytical expressions for the expected photon number, given by Eq.\eqref{npromedio 1}, which calculates the expected photon number for a thermal state. The analytical solution is depicted by the solid blue curve, while the numerical results are shown as dots. The cases considered are: (a) $\gamma = 0.45$ and $\bar{n}_0 = 3$, and (b) $\gamma = 0.6$ and $\bar{n}_0 = 2$.}
\label{termico}
\end{figure}

We consider now the case when the initial condition is a thermal state, which can be expressed as
\begin{equation}\label{eq0160}
\hat{\rho}_{\mathrm{th}}(0)=\frac{1}{\bar{n}_0+1}\sum_{n=0}^\infty\left(\frac{\bar{n}_0}{\bar{n}_0+1}\right)^n\ket{n}\bra{n},
\end{equation}
where $\bar{n}_0$ is the initial average number photons of the thermal state. Substituting the above equation into Eq.~\eqref{numeropromedio} and expanding in Taylor series $e^{\hat{J}t}$, we arrive to
\begin{equation}
\bar{n}(t)=e^{-2\gamma t}\mbox{Tr}\bigg[\hat{a}^\dagger\hat{a}\sum_{k=0}^{\infty}\frac{(\hat{J}t)^k}{k!}\hat{\rho}_{\mathrm{th}}(0)\bigg].
\end{equation}
Writing explicitly the superoperator, we get
\begin{equation}
\bar{n}(t)=e^{-2\gamma t}\mbox{Tr}\Bigg[\sum_{n=0}^\infty\frac{(2\gamma t)^n}{n!}\hat{a}^\dagger\hat{a}\hat{V}^n\hat{\rho}_{\mathrm{th}}(0)\hat{V}^{\dagger n}\Bigg].
\end{equation}
It can be proved that (see Appendix \ref{appendix b})
\begin{equation}
\hat{V}^k\hat{\rho}_{\mathrm{th}}(0)\hat{V}^{\dagger k}=\left(\frac{\bar{n}_0}{\bar{n}_0+1}\right)^k\hat{\rho}_{\mathrm{th}}(0),
\end{equation}
which means that the density matrix for a thermal distribution is an eigendensity matrix of the jump operator $\hat{J}$ with eigenvalue $\frac{\bar{n}_0}{\bar{n}_0+1}$. Upon applying the operator and after some algebra, we may be convinced that
\begin{equation}\label{npromedio 1}
\bar{n}(t)=\exp\left(\frac{-2\gamma t}{\bar{n}_0+1}\right)\bar{n}_0; 
\end{equation}
so, a thermal state decays exponentially under a Lindblad equation whose operators are given by the Susskind-Glogower operators. The explicit evolution of the mean photon number is depicted in Fig.~\ref{termico}. The initial state is the thermal state; we display two different cases in which we can observe the decay of the initial average photon number over time for different values of $\bar{n}_0$ which tells us the starting point, and different $\gamma$ values. As $\gamma$ increases, we observe a more rapid decrease in the average photon number because evolution decays exponentially. Furthermore, for longer times all the interaction is lost in the environment as expected.

\section{Case $[\hat{A},\hat{B}]=2\hat{A}\hat{B}$}\label{2AB}
Given that operators $\hat{A}$ and $\hat{B}$ satisfy the commutation relation $[\hat{A}, \hat{B}] = 2\hat{A}\hat{B}$, or equivalently $\hat{B}\hat{A} = -\hat{A}\hat{B}$, it can be rigorously proven by mathematical induction that for every non-negative integer $n$,
\begin{equation}
\begin{split}
(\omega \hat{A}+ g \hat{B})^{2n}&=(\omega^2 \hat{A}^2+ g^2\hat{B}^2)^n,\\
(\omega\hat{A}+ g \hat{B})^{2n+1} &=(\omega \hat{A}+g \hat{B})(\omega^2\hat{A}^2+ g^2\hat{B}^2)^n;
\end{split}
\end{equation}
then, the exponential operator 
\begin{equation}\label{301}
\begin{split}
\exp\left[-it\left(\omega\hat{A}+g\hat{B}\right)\right]&=\sum_{n=0}^\infty\frac{(-it)^n}{n!}(\omega\hat{A}+g\hat{B})^n\\
&=\sum_{n=0}^\infty\frac{(-it)^{2n}}{(2n)!}(\omega\hat{A}+g\hat{B})^{2n}+\sum_{n=0}^\infty\frac{(-it)^{2n+1}}{(2n+1)!}(\omega\hat{A}+ g \hat{B})^{2n+1}
\end{split}
\end{equation}
can be written as
\begin{equation}
\begin{split}
\exp\left[-it\left(\omega\hat{A}+g\hat{B}\right)\right]&=\sum_{n=0}^\infty\frac{(-it)^{2n}}{(2n)!}(\omega^2\hat{A}^2+ g^2\hat{B}^2)^n\\&+\sum_{n=0}^\infty\frac{(-it)^{2n+1}}{(2n+1)!}(\omega\hat{A}+g\hat{B})(\omega^2\hat{A}^2+ g^2\hat{B}^2)^n,
\end{split}
\end{equation}
and identifying the last two series with the sine and the cosine, we get
\begin{equation}\label{exp2}
\exp\left[-it\left(\omega\hat{A}+ g\hat{B}\right)\right]= \cos\left(t\sqrt{\omega^2\hat{A}^2+ g^2\hat{B}^2}\right)
-i \frac{\omega\hat{A}+ g\hat{B}}{\sqrt{\omega^2\hat{A}^2+ g^2 \hat{B}^2}}\sin\left(t\sqrt{\omega^2\hat{A}^2+g^2\hat{B}^2}\right).   
\end{equation}

As a concrete example of this case, we make the identifications
\begin{equation}
\hat{A} \rightarrow (-1)^{\hat{n}}, \qquad \hat{B} \rightarrow \hat{x}, 
\end{equation}
where $\hat{n}$ is the number operator, so $(-1)^{\hat{n}}$ represents the photon-number parity operator \cite{gerry2023introductory,gerry2010parity}, and $\hat{x}$ is the position operator \cite{sakurai2020modern}. If we express the position operator in terms of the ladder operators as $\hat{x} = (\hat{a} + \hat{a}^\dagger) / \sqrt{2}$, we verify that the commutation relation $[(-1)^{\hat{n}}, \hat{x}] = 2 (-1)^{\hat{n}} \hat{x}$ is satisfied, and this relation can be extended to any well behaved odd function $f_{\text{odd}}(\hat{x})$ of $\hat{x}$.

We considered now a binary Glauber-Fock photonic lattice described by the Hamiltonian, $\hat{H}= \omega (-1)^{\hat{n}} + g \hat{x}$. This Hamiltonian is a semi-infinite tridiagonal matrix, indicating that interactions occur only between nearest neighbors; the off-diagonal elements represent the tunneling rates between sites, following a square-root-law distribution; the diagonal elements reflect changes in the effective refractive index in each waveguide. These type of arrays can be fabricated in a photopolymer using deep UV lithography or in silicon using electron beam lithography, and the required square root law for couplings between guides can be imposed by carefully varying the spacing between waveguides \cite{jordana2007deep,szameit2007control,bojko2011electron}. The system's dynamics is governed by the Schrödinger type equation $i\partial_z|\psi(z)\rangle = \hat{H}|\psi(z)\rangle$, where $z$ is a real parameter which represents the propagation distance of the field; the formal solution of this Schrödinger type equation is
\begin{equation}
|\psi(z)\rangle = \exp\left\{-iz\left[\omega (-1)^{\hat{n}} + g \hat{x}\right]\right\}|\psi(0)\rangle.
\end{equation}
The initial state in the continuous basis is given by
\begin{equation}
|\psi(0)\rangle=\int_{-\infty}^{\infty}\psi(x';0)|x'\rangle \,dx',
\end{equation}
and using Eq.~\eqref{exp2}, the solution can be written as 
\begin{equation}
|\psi(z)\rangle= \biggl[\cos\left(z\sqrt{\omega^2+ g^2 \hat{x}^2}\right)
-i \frac{\omega (-1)^{\hat{n}}+ g \hat{x}}{\sqrt{\omega^2 + g^2 \hat{x}^2}}\sin\left(z\sqrt{\omega^2 +g^2 \hat{x}^2}\right)\biggr]\int_{-\infty}^{\infty}\psi(x';0)|x'\rangle \,dx'.
\end{equation}
In the above equation, we proceed to apply each of the operators. As $f(\hat{x})|x'\rangle= f(x')|x'\rangle$ and $(-1)^{\hat{n}}|x'\rangle=|-x'\rangle$, we have 
\begin{equation}\label{ketpsi}
\begin{split}
|\psi(x;z)\rangle&= \int_{-\infty}^{\infty}\psi(x';0)\biggl[ \cos\left(z\sqrt{\omega^2+ g^2 x^2}\right)-i g \frac{x\sin\left(z\sqrt{\omega^2 +g^2 x^2}\right)}{\sqrt{\omega^2 + g^2 x^2}}\biggr] |x'\rangle\\ &-i \psi(x';0)\omega\frac{\sin\left(z\sqrt{\omega^2 +g^2 x^2}\right)}{\sqrt{\omega^2 + g^2 x^2}}|-x'\rangle \,dx';
\end{split}
\end{equation}
projecting \eqref{ketpsi} with $\bra{x}$, finally we get 
\begin{equation}\label{onda}
 \begin{split}
 \psi(x;z) &=  \biggl[\cos\left(z\sqrt{\omega^2+ g^2 x^2}\right)- ig \frac{x \sin\left(z\sqrt{\omega^2 +g^2 x^2}\right) }{\sqrt{\omega^2 + g^2 x^2}}\biggr]\psi(x;0)\\
&-i\omega\frac{\sin\left(z\sqrt{\omega^2 +g^2 x^2}\right) }{\sqrt{\omega^2 + g^2 x^2}}\psi(-x;0).
\end{split}
\end{equation}
The optical field amplitude along the $m$-th waveguide after a propagation distance $z$ is given by $E_m(z)=\langle m|\psi(z)\rangle$; using the identity decomposition in the position basis, we obtain
\begin{equation}\label{campo}
E_m(z) = \int_{-\infty}^{\infty}\varphi_m(x)\psi(x;z) dx,
\end{equation}
where $\varphi_m(x)$ are the normalized simple harmonic oscillator wave functions and $\psi(x;z)$ is given by \eqref{onda}.\\

\begin{figure}
\centering
\includegraphics[width=\linewidth]{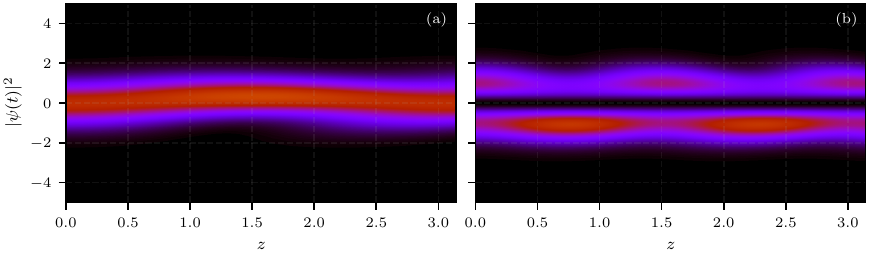}
\caption{The squared absolute value of the field described by Eq.~\eqref{onda} with $\omega=1$ and $g=0.45$ is shown. In panel (a), the initial condition $\psi(x;0)$ is a Gaussian function. Panel (b) depicts $\psi(x;0)$ as a first-order Hermite-Gauss function. These different initial wavefunctions illustrate distinct spatial distributions, from a symmetric Gaussian profile to the more structured pattern of a Hermite-Gauss mode with a single node.}
\label{figonda}
\end{figure}

Fig.~\ref{figonda} presents a contour plot of the intensity dynamics, depicting the propagation as a function of distance $z$ according to Eq.~\eqref{onda}. The initial states used are: (a) the Gaussian function $\psi(x;0) = \pi^{-1/4}\mathrm{e}^{-x^2/2}$, and (b) the first-order Hermite-Gauss function $\psi(x;0) = \frac{\pi^{-1/4}}{\sqrt{2^n n!}}\mathrm{e}^{-x^2/2}H_n(x)$. In panel (a), the plot shows a central intensity band predominantly located at $x=0$. In panel (b), two nearly symmetrical intensity bands relative to $x=0$ are observed, which become more pronounced as the order $n$ increases.

\begin{figure}
\includegraphics[width=\linewidth]{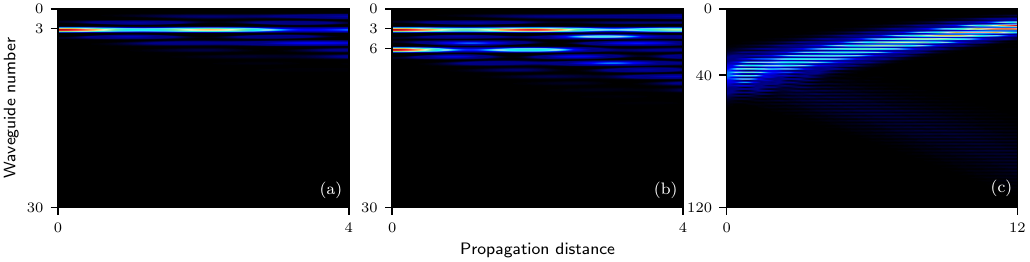}
\caption{Examining the intensity evolution among waveguide sites in this Glauber-Fock lattice with $\omega=1$ and $g=0.5$: panel (a) shows the excitation of the third waveguide site, highlighting localized intensity changes. Panel (b) illustrates simultaneous illumination of waveguides 3 and 6, indicating interactions between neighboring sites. Panel (c) demonstrates the evolution of a coherent state with an average photon number $\langle\hat{n}\rangle=40$, revealing the influence of global coherence on intensity distribution across the lattice.}
\label{guias}
\end{figure}

Fig.~\ref{guias} depicts the squared absolute value of the field $E_m(z)$, as given by Eq.~\eqref{campo}. Panel (a) presents the initial condition as a Hermite-Gauss function of order $n$, $\psi(x;0) = \varphi_n(x)$, where $n$ corresponds to the $n$-th illuminated waveguide. Panel (b) shows the initial condition as a superposition of two Hermite-Gauss functions of orders $j$ and $k$, $\psi(x;0) = \varphi_j(x) + \varphi_k(x)$, with $j$ and $k$ indicating the illuminated waveguides. Panel (c) illustrates the initial condition as a coherent wave function, $\psi(x;0) = \psi_\alpha(x)$. The figures reveal that, as light propagates through the waveguides, the energy does not quickly spread from top to bottom, which is a common behavior in other Glauber-Fock photonic lattices. Instead, a greater propagation distance is required to observe this shift.

\section{Conclusions}\label{conclusiones}
It has been demonstrated that certain exponential functions involving the sum of two non-commuting operators can be factorized, despite not conforming to conventional algebraic closures. This factorization method holds significant implications, particularly in quantum mechanics. Specifically, it has proven effective in the context of Lindblad operators, governing systems where only one excitation can leak out, resulting in observed exponential decay in thermal density matrices. Moreover, this method has been generalized to handle arbitrary initial density matrices, thereby extending its applicability across diverse quantum states and scenarios.

In addition to addressing non-commuting operators, we have investigated the factorization of exponential functions under conditions where the operators commute, specifically when $[\hat{A},\hat{B}] = 2\hat{A}\hat{B}$. This case is particularly valuable for analyzing the propagation dynamics of classical electromagnetic fields within Glauber-Fock waveguide arrays, crucial components in photonics for manipulating quantum-level light. By employing this factorization method, we can gain deeper insights into the complex dynamics and interactions of electromagnetic fields within these structured environments.

\appendix
\section{}\label{appendix a}
In this appendix we show the evolution of $\hat{\rho}(t)$ when we consider a coherent state $\hat{\rho}(0)=|\alpha\rangle\langle \alpha|$ as initial condition. From \eqref{solutionf}, we get that
\begin{align}
e^{2\gamma t}\hat{\rho} t)&=e^{\hat{J}t}\ket{\alpha}\bra{\alpha}+\ket{0}\bra{0}\sum_{k=m}^\infty\frac{\gamma^k t^k}{k!}\sum_{m=1}^k2^{k-m}(2^m-2)\hat{V}^{(k-m)}\ket{\alpha}\bra{\alpha}\hat{V}^{\dagger(k-m)}\ket{0}\bra{0}
\nonumber\\
&+\ket{0}\bra{0}\sum_{k=m}^\infty\frac{\gamma^k t^k}{k!}\sum_{m=1}^k2^{(k-m)}\hat{V}^{(k-m)}\ket{\alpha}\bra{\alpha}\hat{V}^{\dagger(k-m)}
\nonumber\\
&+\sum_{k=m}^\infty\frac{\gamma^k t^k}{k!}\sum_{m=1}^k2^{(k-m)}\hat{V}^{(k-m)}\ket{\alpha}\bra{\alpha}\hat{V}^{\dagger(k-m)}\ket{0}\bra{0}.
\end{align}
We can expand the first term in Taylor series, use the fact that $\hat{J}^n\hat{\rho}(0)=\hat{V}^n\hat{\rho}(0)\hat{V}^{\dagger n}$, and that $\hat{V}^{\dagger n}\ket{0}=\ket{n}$ to obtain
\begin{align}
e^{2\gamma t}\hat{\rho}(t) &= \sum_{n=0}^{\infty}\frac{(2\gamma t)^n}{n!}\hat{V}^n\ket{\alpha}\bra{\alpha}\hat{V}^{\dagger n}\nonumber\\ &+ \ket{0}\bra{0}\sum_{k=m}^\infty\frac{\gamma^k t^k}{k!}\sum_{m=1}^k2^{k-m}(2^m-2)\bra{k-m}\alpha\rangle\langle \alpha\ket{k-m}
\nonumber \\
& +\sum_{k=m}^{\infty}\frac{(2\gamma t)^k}{k!}\sum_{m=1}^{k}2^{k-m}\ket{0}\bra{k-m}\alpha\rangle\bra{\alpha}\hat{V}^{\dagger(k-m)}\nonumber\\ &+ \sum_{k=m}^{\infty}\frac{(2\gamma t)^k}{k!}\sum_{m=1}^{k}2^{k-m}\hat{V}^{k-m}\ket{\alpha}\bra{\alpha}k-m\rangle\ket{0}.
\end{align}
Writing $\ket{\alpha}$ as a superposition of Fock states $\ket{\alpha}=e^{\frac{-|\alpha|^2}{2}}\sum_{j=0}^{\infty}\frac{\alpha^j}{\sqrt{j!}}\ket{j} $, we can prove that
\begin{equation}\label{acoherent}
\bra{k-m}\alpha\rangle= e^{\frac{-|\alpha|^2}{2}}\frac{\alpha^{k-m}}{\sqrt{(k-m)!}}, \qquad \hat{V}^n\ket{\alpha}= e^{\frac{-|\alpha|^2}{2}}\sum_{j=0}^{\infty}\frac{\alpha^{j+n}}{\sqrt{(j+n)!}}\ket{j}, 
\end{equation}
as well as for its adjoints, and then we arrive to
\begin{align}
e^{2\gamma t}\hat{\rho}(t) &= \sum_{n=0}^{\infty}\frac{(2|\alpha|\gamma t)^n}{n!}e^{-|\alpha|^2}\sum_{j,j'=0}^{\infty}\frac{\alpha^j\alpha^{j'}}{\sqrt{(j+n)!(j'+n)!}}\ket{j'}\bra{j}\nonumber\\&+e^{-|\alpha|^2}\ket{0}\bra{0}\sum_{k=m}^\infty\frac{\gamma^k t^k}{k!}\sum_{m=1}^k2^{k-m}(2^m-2)\frac{|\alpha|^{2(k-m)}}{(k-m)!}
\nonumber \\
& +e^{-|\alpha|^2}\sum_{k=m}^\infty\frac{\gamma^k t^k}{k!}\sum_{m=1}^k2^{k-m}\frac{|\alpha|^{2(k-m)}}{\sqrt{(k-m)!}}\sum_{j=0}^{\infty}\frac{\alpha^{*j}}{\sqrt{(j+k-m)!}}\ket{0}\bra{j}
\nonumber \\
&+e^{-|\alpha|^2}\sum_{k=m}^\infty\frac{\gamma^k t^k}{k!}\sum_{m=1}^k2^{k-m}\frac{|\alpha|^{2(k-m)}}{\sqrt{(k-m)!}}\sum_{j=0}^{\infty}\frac{\alpha^{j}}{\sqrt{(j+k-m)!}}\ket{j}\bra{0},
\end{align}
We extend the sum over $m$ to infinity because we are only adding zeros as we have $(k-m)!$ in the denominator, and changing the order of the sums, it is easy to see that
\begin{align}
e^{2\gamma t}\hat{\rho}(t) &= e^{-|\alpha|^2}\sum_{n=0}^{\infty}\frac{(2|\alpha|\gamma t)^n}{n!}\sum_{j,j'=0}^{\infty}\frac{\alpha^j\alpha^{j'}}{\sqrt{(j+n)!(j'+n)!}}\ket{j'}\bra{j}\nonumber\\&+ e^{-|\alpha|^2}\ket{0}\bra{0}\sum_{m=1}^{\infty}\sum_{k=m}^\infty\frac{\gamma^k t^k}{k!}2^{k-m}(2^m-2)\frac{|\alpha|^{2(k-m)}}{(k-m)!}
\nonumber \\
& +e^{-|\alpha|^2}\sum_{m=1}^{\infty}\sum_{k=m}^\infty\frac{\gamma^k t^k}{k!}2^{k-m}\frac{|\alpha|^{2(k-m)}}{\sqrt{(k-m)!}}\sum_{j=0}^{\infty}\frac{\alpha^{*j}}{\sqrt{(j+k-m)!}}\ket{0}\bra{j}
\nonumber \\
&+e^{-|\alpha|^2}\sum_{m=1}^{\infty}\sum_{k=m}^\infty\frac{\gamma^k t^k}{k!}2^{k-m}\frac{|\alpha|^{2(k-m)}}{\sqrt{(k-m)!}}\sum_{j=0}^{\infty}\frac{\alpha^{j}}{\sqrt{(j+k-m)!}}\ket{j}\bra{0}.
\end{align}
Finally, taking $n=k-m$ we obtain Eq.~\eqref{solcoherente}. 

\section{}\label{appendix b}
In this appendix, we show that
\begin{equation}\label{appen 1}
\hat{V}^k\hat{\rho}_{\mathrm{th}}(0)\hat{V}^{\dagger k}=\bigg[\frac{\bar{n}_0}{\bar{n}_0+1}\bigg]^k\hat{\rho}_{\mathrm{th}}(0),
\end{equation}
where $\hat{\rho}_{\mathrm{th}}(0)$ is a thermal state.\\
From Eq. \eqref{eq0160}, we have that 
\begin{equation}
\hat{V}^k\hat{\rho}_{\mathrm{th}}(0)\hat{V}^{\dagger k} =\frac{1}{\bar{n}_0+1}\sum_{n=0}^\infty\bigg[\frac{\bar{n}_0}{\bar{n}_0+1}\bigg]^n\hat{V}^k\ket{n}\bra{n}\hat{V}^{\dagger k};
\end{equation}
as $\hat{V}^k\ket{n}=\ket{n-k}$, we can write 
\begin{equation}
\hat{V}^k\hat{\rho}_{\mathrm{th}}(0)\hat{V}^{\dagger k} =\frac{1}{\bar{n}_0+1}\sum_{n=k}^\infty\bigg[\frac{\bar{n}_0}{\bar{n}_0+1}\bigg]^n\ket{n-k}\bra{n-k}.
\end{equation}
Taking $m=n-k$, and groping terms 
\begin{equation}
\hat{V}^k\hat{\rho}_{\mathrm{th}}(0)\hat{V}^{\dagger k} = \bigg[\frac{\bar{n}_0}{\bar{n}_0+1}\bigg]^k \frac{1}{\bar{n}_0+1}\sum_{m=0}^\infty\bigg[\frac{\bar{n}_0}{\bar{n}_0+1}\bigg]^m\ket{m}\bra{m}.
\end{equation}
Finally, we can easily identify the thermal state on the right-hand side of the above equation to obtain \eqref{appen 1}, as desired.


\begin{thebibliography}{53}%
\makeatletter
\providecommand \@ifxundefined [1]{%
 \@ifx{#1\undefined}
}%
\providecommand \@ifnum [1]{%
 \ifnum #1\expandafter \@firstoftwo
 \else \expandafter \@secondoftwo
 \fi
}%
\providecommand \@ifx [1]{%
 \ifx #1\expandafter \@firstoftwo
 \else \expandafter \@secondoftwo
 \fi
}%
\providecommand \natexlab [1]{#1}%
\providecommand \enquote  [1]{``#1''}%
\providecommand \bibnamefont  [1]{#1}%
\providecommand \bibfnamefont [1]{#1}%
\providecommand \citenamefont [1]{#1}%
\providecommand \href@noop [0]{\@secondoftwo}%
\providecommand \href [0]{\begingroup \@sanitize@url \@href}%
\providecommand \@href[1]{\@@startlink{#1}\@@href}%
\providecommand \@@href[1]{\endgroup#1\@@endlink}%
\providecommand \@sanitize@url [0]{\catcode `\\12\catcode `\$12\catcode
  `\&12\catcode `\#12\catcode `\^12\catcode `\_12\catcode `\%12\relax}%
\providecommand \@@startlink[1]{}%
\providecommand \@@endlink[0]{}%
\providecommand \url  [0]{\begingroup\@sanitize@url \@url }%
\providecommand \@url [1]{\endgroup\@href {#1}{\urlprefix }}%
\providecommand \urlprefix  [0]{URL }%
\providecommand \Eprint [0]{\href }%
\providecommand \doibase [0]{http://dx.doi.org/}%
\providecommand \selectlanguage [0]{\@gobble}%
\providecommand \bibinfo  [0]{\@secondoftwo}%
\providecommand \bibfield  [0]{\@secondoftwo}%
\providecommand \translation [1]{[#1]}%
\providecommand \BibitemOpen [0]{}%
\providecommand \bibitemStop [0]{}%
\providecommand \bibitemNoStop [0]{.\EOS\space}%
\providecommand \EOS [0]{\spacefactor3000\relax}%
\providecommand \BibitemShut  [1]{\csname bibitem#1\endcsname}%
\let\auto@bib@innerbib\@empty
\bibitem [{\citenamefont {Yoshida}(1993)}]{yoshida1993recent}%
  \BibitemOpen
  \bibfield  {author} {\bibinfo {author} {\bibfnamefont {H.}~\bibnamefont
  {Yoshida}},\ }in\ \href {\doibase
  https://doi.org/10.1007/978-94-011-2030-2_3} {\emph {\bibinfo {booktitle}
  {Qualitative and Quantitative Behaviour of Planetary Systems: Proceedings of
  the Third Alexander von Humboldt Colloquium on Celestial Mechanics}}}\
  (\bibinfo {organization} {Springer},\ \bibinfo {year} {1993})\ pp.\ \bibinfo
  {pages} {27--43}\BibitemShut {NoStop}%
\bibitem [{\citenamefont {McLachlan}\ and\ \citenamefont
  {Quispel}(2002)}]{mclachlan2002splitting}%
  \BibitemOpen
  \bibfield  {author} {\bibinfo {author} {\bibfnamefont {R.~I.}\ \bibnamefont
  {McLachlan}}\ and\ \bibinfo {author} {\bibfnamefont {G.~R.~W.}\ \bibnamefont
  {Quispel}},\ }\href {\doibase 10.1017/S0962492902000053} {\bibfield
  {journal} {\bibinfo  {journal} {Acta Numerica}\ }\textbf {\bibinfo {volume}
  {11}},\ \bibinfo {pages} {341–434} (\bibinfo {year} {2002})}\BibitemShut
  {NoStop}%
\bibitem [{\citenamefont {Plass}(1961)}]{plass1961classical}%
  \BibitemOpen
  \bibfield  {author} {\bibinfo {author} {\bibfnamefont {G.~N.}\ \bibnamefont
  {Plass}},\ }\href {\doibase 10.1103/RevModPhys.33.37} {\bibfield  {journal}
  {\bibinfo  {journal} {Rev. Mod. Phys.}\ }\textbf {\bibinfo {volume} {33}},\
  \bibinfo {pages} {37} (\bibinfo {year} {1961})}\BibitemShut {NoStop}%
\bibitem [{\citenamefont {Tapia-Valerdi}\ \emph {et~al.}(2024)\citenamefont
  {Tapia-Valerdi}, \citenamefont {Ramos-Prieto}, \citenamefont {Soto-Eguibar},\
  and\ \citenamefont {Moya-Cessa}}]{tapia2024generalization}%
  \BibitemOpen
  \bibfield  {author} {\bibinfo {author} {\bibfnamefont {M.~A.}\ \bibnamefont
  {Tapia-Valerdi}}, \bibinfo {author} {\bibfnamefont {I.}~\bibnamefont
  {Ramos-Prieto}}, \bibinfo {author} {\bibfnamefont {F.}~\bibnamefont
  {Soto-Eguibar}}, \ and\ \bibinfo {author} {\bibfnamefont {H.~M.}\
  \bibnamefont {Moya-Cessa}},\ }\href {\doibase 10.18576/amis/180220}
  {\bibfield  {journal} {\bibinfo  {journal} {Appl. Math}\ }\textbf {\bibinfo
  {volume} {18}},\ \bibinfo {pages} {463} (\bibinfo {year} {2024})}\BibitemShut
  {NoStop}%
\bibitem [{\citenamefont {Yu}\ and\ \citenamefont
  {Tong}(2023)}]{yu2023evolution}%
  \BibitemOpen
  \bibfield  {author} {\bibinfo {author} {\bibfnamefont {X.-D.}\ \bibnamefont
  {Yu}}\ and\ \bibinfo {author} {\bibfnamefont {D.~M.}\ \bibnamefont {Tong}},\
  }\href {\doibase 10.1103/PhysRevLett.131.200202} {\bibfield  {journal}
  {\bibinfo  {journal} {Phys. Rev. Lett.}\ }\textbf {\bibinfo {volume} {131}},\
  \bibinfo {pages} {200202} (\bibinfo {year} {2023})}\BibitemShut {NoStop}%
\bibitem [{\citenamefont {Teuber}\ and\ \citenamefont
  {Scheel}(2020)}]{teuber2020solving}%
  \BibitemOpen
  \bibfield  {author} {\bibinfo {author} {\bibfnamefont {L.}~\bibnamefont
  {Teuber}}\ and\ \bibinfo {author} {\bibfnamefont {S.}~\bibnamefont
  {Scheel}},\ }\href {\doibase 10.1103/PhysRevA.101.042124} {\bibfield
  {journal} {\bibinfo  {journal} {Phys. Rev. A}\ }\textbf {\bibinfo {volume}
  {101}},\ \bibinfo {pages} {042124} (\bibinfo {year} {2020})}\BibitemShut
  {NoStop}%
\bibitem [{\citenamefont {Sack}(1958)}]{sack1958taylor}%
  \BibitemOpen
  \bibfield  {author} {\bibinfo {author} {\bibfnamefont {R.~A.}\ \bibnamefont
  {Sack}},\ }\href {\doibase 10.1080/14786435808244572} {\bibfield  {journal}
  {\bibinfo  {journal} {Phil. Mag.,}\ }\textbf {\bibinfo {volume} {3}},\
  \bibinfo {pages} {497} (\bibinfo {year} {1958})}\BibitemShut {NoStop}%
\bibitem [{\citenamefont {Wei}\ and\ \citenamefont
  {Norman}(1963)}]{wei1963lie}%
  \BibitemOpen
  \bibfield  {author} {\bibinfo {author} {\bibfnamefont {J.}~\bibnamefont
  {Wei}}\ and\ \bibinfo {author} {\bibfnamefont {E.}~\bibnamefont {Norman}},\
  }\href {\doibase 10.1063/1.1703993} {\bibfield  {journal} {\bibinfo
  {journal} {J. Math. Phys.}\ }\textbf {\bibinfo {volume} {4}},\ \bibinfo
  {pages} {575} (\bibinfo {year} {1963})}\BibitemShut {NoStop}%
\bibitem [{\citenamefont {Wilcox}(1967)}]{wilcox1967exponential}%
  \BibitemOpen
  \bibfield  {author} {\bibinfo {author} {\bibfnamefont {R.~M.}\ \bibnamefont
  {Wilcox}},\ }\href {\doibase 10.1063/1.1705306} {\bibfield  {journal}
  {\bibinfo  {journal} {J. Math. Phys.}\ }\textbf {\bibinfo {volume} {8}},\
  \bibinfo {pages} {962} (\bibinfo {year} {1967})}\BibitemShut {NoStop}%
\bibitem [{\citenamefont {Suzuki}(1985)}]{suzuki1985decomposition}%
  \BibitemOpen
  \bibfield  {author} {\bibinfo {author} {\bibfnamefont {M.}~\bibnamefont
  {Suzuki}},\ }\href {\doibase 10.1063/1.526596} {\bibfield  {journal}
  {\bibinfo  {journal} {J. Math. Phys.}\ }\textbf {\bibinfo {volume} {26}},\
  \bibinfo {pages} {601} (\bibinfo {year} {1985})}\BibitemShut {NoStop}%
\bibitem [{\citenamefont {Ban}(1993)}]{ban1993decomposition}%
  \BibitemOpen
  \bibfield  {author} {\bibinfo {author} {\bibfnamefont {M.}~\bibnamefont
  {Ban}},\ }\href {\doibase 10.1364/JOSAB.10.001347} {\bibfield  {journal}
  {\bibinfo  {journal} {J. Opt. Soc. Am. B}\ }\textbf {\bibinfo {volume}
  {10}},\ \bibinfo {pages} {1347} (\bibinfo {year} {1993})}\BibitemShut
  {NoStop}%
\bibitem [{\citenamefont {Louisell}(1973)}]{louisell}%
  \BibitemOpen
  \bibfield  {author} {\bibinfo {author} {\bibfnamefont {W.~H.}\ \bibnamefont
  {Louisell}},\ }\href {https://www.osti.gov/biblio/4208943} {\emph {\bibinfo
  {title} {Quantum statistical properties of radiation}}}\ (\bibinfo {year}
  {1973})\BibitemShut {NoStop}%
\bibitem [{\citenamefont {Rossmann}(2002)}]{RossmannW}%
  \BibitemOpen
  \bibfield  {author} {\bibinfo {author} {\bibfnamefont {W.}~\bibnamefont
  {Rossmann}},\ }\href
  {https://books.google.com/books/about/Lie_Groups.html?id=EjDazZvcquwC} {\emph
  {\bibinfo {title} {Lie Groups: An Introduction Through Linear Groups}}}\
  (\bibinfo  {publisher} {Oxford University Press},\ \bibinfo {year}
  {2002})\BibitemShut {NoStop}%
\bibitem [{\citenamefont {Hall}(2013)}]{Hall_2013}%
  \BibitemOpen
  \bibfield  {author} {\bibinfo {author} {\bibfnamefont {B.~C.}\ \bibnamefont
  {Hall}},\ }\href {\doibase 10.1007/978-1-4614-7116-5_16} {\emph {\bibinfo
  {title} {Lie Groups, Lie Algebras, and Representations}}}\ (\bibinfo
  {publisher} {Springer New York},\ \bibinfo {year} {2013})\BibitemShut
  {NoStop}%
\bibitem [{\citenamefont {Breuer}\ and\ \citenamefont
  {Petruccione}(2002)}]{breuer2002theory}%
  \BibitemOpen
  \bibfield  {author} {\bibinfo {author} {\bibfnamefont {H.}~\bibnamefont
  {Breuer}}\ and\ \bibinfo {author} {\bibfnamefont {F.}~\bibnamefont
  {Petruccione}},\ }\href {https://books.google.com.mx/books?id=0Yx5VzaMYm8C}
  {\emph {\bibinfo {title} {The Theory of Open Quantum Systems}}}\ (\bibinfo
  {publisher} {Oxford University Press},\ \bibinfo {year} {2002})\BibitemShut
  {NoStop}%
\bibitem [{\citenamefont {Carmichael}(2009)}]{carmichael1993open}%
  \BibitemOpen
  \bibfield  {author} {\bibinfo {author} {\bibfnamefont {H.}~\bibnamefont
  {Carmichael}},\ }\href {https://books.google.com.mx/books?id=uor_CAAAQBAJ}
  {\enquote {\bibinfo {title} {An open systems approach to quantum optics:
  Lectures presented at the universit{\'e} libre de bruxelles, october 28 to
  november 4, 1991},}\ } (\bibinfo {year} {2009})\BibitemShut {NoStop}%
\bibitem [{\citenamefont {Manzano}(2020)}]{manzano2020short}%
  \BibitemOpen
  \bibfield  {author} {\bibinfo {author} {\bibfnamefont {D.}~\bibnamefont
  {Manzano}},\ }\href {\doibase 10.1063/1.5115323} {\bibfield  {journal}
  {\bibinfo  {journal} {AIP Advances}\ }\textbf {\bibinfo {volume} {10}},\
  \bibinfo {pages} {025106} (\bibinfo {year} {2020})}\BibitemShut {NoStop}%
\bibitem [{\citenamefont {Risken}(1996)}]{risken1984solutions}%
  \BibitemOpen
  \bibfield  {author} {\bibinfo {author} {\bibfnamefont {H.}~\bibnamefont
  {Risken}},\ }\bibfield  {booktitle} {\emph {\bibinfo {booktitle} {The
  Fokker-Planck Equation: Methods of Solution and Applications}},\ }\href
  {\doibase 10.1007/978-3-642-61544-3_4} {\ ,\ \bibinfo {pages} {63} (\bibinfo
  {year} {1996})}\BibitemShut {NoStop}%
\bibitem [{\citenamefont {Gardiner}\ and\ \citenamefont
  {Zoller}(2004)}]{gardiner1985handbook}%
  \BibitemOpen
  \bibfield  {author} {\bibinfo {author} {\bibfnamefont {C.}~\bibnamefont
  {Gardiner}}\ and\ \bibinfo {author} {\bibfnamefont {P.}~\bibnamefont
  {Zoller}},\ }\href {https://books.google.com.mx/books?id=a_xsT8oGhdgC} {\emph
  {\bibinfo {title} {Quantum Noise: A Handbook of Markovian and Non-Markovian
  Quantum Stochastic Methods with Applications to Quantum Optics}}},\ Springer
  Series in Synergetics\ (\bibinfo  {publisher} {Springer},\ \bibinfo {year}
  {2004})\BibitemShut {NoStop}%
\bibitem [{\citenamefont {Yuto~Ashida}\ and\ \citenamefont
  {Ueda}(2020)}]{ashida2020non}%
  \BibitemOpen
  \bibfield  {author} {\bibinfo {author} {\bibfnamefont {Z.~G.}\ \bibnamefont
  {Yuto~Ashida}}\ and\ \bibinfo {author} {\bibfnamefont {M.}~\bibnamefont
  {Ueda}},\ }\href {\doibase 10.1080/00018732.2021.1876991} {\bibfield
  {journal} {\bibinfo  {journal} {Advances in Physics}\ }\textbf {\bibinfo
  {volume} {69}},\ \bibinfo {pages} {249} (\bibinfo {year} {2020})}\BibitemShut
  {NoStop}%
\bibitem [{\citenamefont {Long}\ \emph {et~al.}(2022)\citenamefont {Long},
  \citenamefont {Xue},\ and\ \citenamefont {Zhang}}]{long2022non}%
  \BibitemOpen
  \bibfield  {author} {\bibinfo {author} {\bibfnamefont {Y.}~\bibnamefont
  {Long}}, \bibinfo {author} {\bibfnamefont {H.}~\bibnamefont {Xue}}, \ and\
  \bibinfo {author} {\bibfnamefont {B.}~\bibnamefont {Zhang}},\ }\href
  {\doibase 10.1103/PhysRevB.105.L100102} {\bibfield  {journal} {\bibinfo
  {journal} {Phys. Rev. B}\ }\textbf {\bibinfo {volume} {105}},\ \bibinfo
  {pages} {L100102} (\bibinfo {year} {2022})}\BibitemShut {NoStop}%
\bibitem [{\citenamefont {Bender}\ and\ \citenamefont
  {Boettcher}(1998)}]{bender1998real}%
  \BibitemOpen
  \bibfield  {author} {\bibinfo {author} {\bibfnamefont {C.~M.}\ \bibnamefont
  {Bender}}\ and\ \bibinfo {author} {\bibfnamefont {S.}~\bibnamefont
  {Boettcher}},\ }\href {\doibase 10.1103/PhysRevLett.80.5243} {\bibfield
  {journal} {\bibinfo  {journal} {Phys. Rev. Lett.}\ }\textbf {\bibinfo
  {volume} {80}},\ \bibinfo {pages} {5243} (\bibinfo {year}
  {1998})}\BibitemShut {NoStop}%
\bibitem [{\citenamefont {Hern\'{a}ndez-S\'{a}nchez}\ \emph
  {et~al.}(2023)\citenamefont {Hern\'{a}ndez-S\'{a}nchez}, \citenamefont
  {Ramos-Prieto}, \citenamefont {Soto-Eguibar},\ and\ \citenamefont
  {Moya-Cessa}}]{hernandez2023exact}%
  \BibitemOpen
  \bibfield  {author} {\bibinfo {author} {\bibfnamefont {L.}~\bibnamefont
  {Hern\'{a}ndez-S\'{a}nchez}}, \bibinfo {author} {\bibfnamefont
  {I.}~\bibnamefont {Ramos-Prieto}}, \bibinfo {author} {\bibfnamefont
  {F.}~\bibnamefont {Soto-Eguibar}}, \ and\ \bibinfo {author} {\bibfnamefont
  {H.~M.}\ \bibnamefont {Moya-Cessa}},\ }\href {\doibase 10.1364/OL.503837}
  {\bibfield  {journal} {\bibinfo  {journal} {Opt. Lett.}\ }\textbf {\bibinfo
  {volume} {48}},\ \bibinfo {pages} {5435} (\bibinfo {year}
  {2023})}\BibitemShut {NoStop}%
\bibitem [{\citenamefont {Longhi}(2020)}]{longhi2020quantum}%
  \BibitemOpen
  \bibfield  {author} {\bibinfo {author} {\bibfnamefont {S.}~\bibnamefont
  {Longhi}},\ }\href {\doibase 10.1364/OL.386232} {\bibfield  {journal}
  {\bibinfo  {journal} {Opt. Lett.}\ }\textbf {\bibinfo {volume} {45}},\
  \bibinfo {pages} {1591} (\bibinfo {year} {2020})}\BibitemShut {NoStop}%
\bibitem [{\citenamefont {Liu}\ \emph {et~al.}(2021)\citenamefont {Liu},
  \citenamefont {Shao}, \citenamefont {Ma}, \citenamefont {Zhang},
  \citenamefont {You}, \citenamefont {Wu}, \citenamefont {Xiang}, \citenamefont
  {Cui},\ and\ \citenamefont {Zhang}}]{liu2021non}%
  \BibitemOpen
  \bibfield  {author} {\bibinfo {author} {\bibfnamefont {S.}~\bibnamefont
  {Liu}}, \bibinfo {author} {\bibfnamefont {R.}~\bibnamefont {Shao}}, \bibinfo
  {author} {\bibfnamefont {S.}~\bibnamefont {Ma}}, \bibinfo {author}
  {\bibfnamefont {L.}~\bibnamefont {Zhang}}, \bibinfo {author} {\bibfnamefont
  {O.}~\bibnamefont {You}}, \bibinfo {author} {\bibfnamefont {H.}~\bibnamefont
  {Wu}}, \bibinfo {author} {\bibfnamefont {Y.~J.}\ \bibnamefont {Xiang}},
  \bibinfo {author} {\bibfnamefont {T.~J.}\ \bibnamefont {Cui}}, \ and\
  \bibinfo {author} {\bibfnamefont {S.}~\bibnamefont {Zhang}},\ }\href
  {\doibase 10.34133/2021/5608038} {\bibfield  {journal} {\bibinfo  {journal}
  {Research}\ }\textbf {\bibinfo {volume} {2021}} (\bibinfo {year} {2021}),\
  10.34133/2021/5608038}\BibitemShut {NoStop}%
\bibitem [{\citenamefont {Liu}\ \emph {et~al.}(2023)\citenamefont {Liu},
  \citenamefont {Li}, \citenamefont {Yang}, \citenamefont {Shen}, \citenamefont
  {Yang}, \citenamefont {Hang},\ and\ \citenamefont
  {Ezawa}}]{liu2023experimental}%
  \BibitemOpen
  \bibfield  {author} {\bibinfo {author} {\bibfnamefont {B.}~\bibnamefont
  {Liu}}, \bibinfo {author} {\bibfnamefont {Y.}~\bibnamefont {Li}}, \bibinfo
  {author} {\bibfnamefont {B.}~\bibnamefont {Yang}}, \bibinfo {author}
  {\bibfnamefont {X.}~\bibnamefont {Shen}}, \bibinfo {author} {\bibfnamefont
  {Y.}~\bibnamefont {Yang}}, \bibinfo {author} {\bibfnamefont {Z.~H.}\
  \bibnamefont {Hang}}, \ and\ \bibinfo {author} {\bibfnamefont
  {M.}~\bibnamefont {Ezawa}},\ }\href {\doibase
  10.1103/PhysRevResearch.5.043034} {\bibfield  {journal} {\bibinfo  {journal}
  {Phys. Rev. Res.}\ }\textbf {\bibinfo {volume} {5}},\ \bibinfo {pages}
  {043034} (\bibinfo {year} {2023})}\BibitemShut {NoStop}%
\bibitem [{\citenamefont {Minganti}\ \emph {et~al.}(2019)\citenamefont
  {Minganti}, \citenamefont {Miranowicz}, \citenamefont {Chhajlany},\ and\
  \citenamefont {Nori}}]{minganti2019quantum}%
  \BibitemOpen
  \bibfield  {author} {\bibinfo {author} {\bibfnamefont {F.}~\bibnamefont
  {Minganti}}, \bibinfo {author} {\bibfnamefont {A.}~\bibnamefont
  {Miranowicz}}, \bibinfo {author} {\bibfnamefont {R.~W.}\ \bibnamefont
  {Chhajlany}}, \ and\ \bibinfo {author} {\bibfnamefont {F.}~\bibnamefont
  {Nori}},\ }\href {\doibase 10.1103/PhysRevA.100.062131} {\bibfield  {journal}
  {\bibinfo  {journal} {Phys. Rev. A}\ }\textbf {\bibinfo {volume} {100}},\
  \bibinfo {pages} {062131} (\bibinfo {year} {2019})}\BibitemShut {NoStop}%
\bibitem [{\citenamefont {Wiersig}(2020)}]{wiersig2020review}%
  \BibitemOpen
  \bibfield  {author} {\bibinfo {author} {\bibfnamefont {J.}~\bibnamefont
  {Wiersig}},\ }\href {\doibase 10.1364/PRJ.396115} {\bibfield  {journal}
  {\bibinfo  {journal} {Photon. Res.}\ }\textbf {\bibinfo {volume} {8}},\
  \bibinfo {pages} {1457} (\bibinfo {year} {2020})}\BibitemShut {NoStop}%
\bibitem [{\citenamefont {Perez-Leija}\ \emph {et~al.}(2010)\citenamefont
  {Perez-Leija}, \citenamefont {Moya-Cessa}, \citenamefont {Szameit},\ and\
  \citenamefont {Christodoulides}}]{perez2010glauber}%
  \BibitemOpen
  \bibfield  {author} {\bibinfo {author} {\bibfnamefont {A.}~\bibnamefont
  {Perez-Leija}}, \bibinfo {author} {\bibfnamefont {H.}~\bibnamefont
  {Moya-Cessa}}, \bibinfo {author} {\bibfnamefont {A.}~\bibnamefont {Szameit}},
  \ and\ \bibinfo {author} {\bibfnamefont {D.~N.}\ \bibnamefont
  {Christodoulides}},\ }\href {\doibase 10.1364/OL.35.002409} {\bibfield
  {journal} {\bibinfo  {journal} {Opt. Lett.}\ }\textbf {\bibinfo {volume}
  {35}},\ \bibinfo {pages} {2409} (\bibinfo {year} {2010})}\BibitemShut
  {NoStop}%
\bibitem [{\citenamefont {Vicencio}\ \emph {et~al.}(2015)\citenamefont
  {Vicencio}, \citenamefont {Cantillano}, \citenamefont {Morales-Inostroza},
  \citenamefont {Real}, \citenamefont {Mej\'{\i}a-Cort\'es}, \citenamefont
  {Weimann}, \citenamefont {Szameit},\ and\ \citenamefont
  {Molina}}]{vicencio2015observation}%
  \BibitemOpen
  \bibfield  {author} {\bibinfo {author} {\bibfnamefont {R.~A.}\ \bibnamefont
  {Vicencio}}, \bibinfo {author} {\bibfnamefont {C.}~\bibnamefont
  {Cantillano}}, \bibinfo {author} {\bibfnamefont {L.}~\bibnamefont
  {Morales-Inostroza}}, \bibinfo {author} {\bibfnamefont {B.}~\bibnamefont
  {Real}}, \bibinfo {author} {\bibfnamefont {C.}~\bibnamefont
  {Mej\'{\i}a-Cort\'es}}, \bibinfo {author} {\bibfnamefont {S.}~\bibnamefont
  {Weimann}}, \bibinfo {author} {\bibfnamefont {A.}~\bibnamefont {Szameit}}, \
  and\ \bibinfo {author} {\bibfnamefont {M.~I.}\ \bibnamefont {Molina}},\
  }\href {\doibase 10.1103/PhysRevLett.114.245503} {\bibfield  {journal}
  {\bibinfo  {journal} {Phys. Rev. Lett.}\ }\textbf {\bibinfo {volume} {114}},\
  \bibinfo {pages} {245503} (\bibinfo {year} {2015})}\BibitemShut {NoStop}%
\bibitem [{\citenamefont {Trompeter}\ \emph {et~al.}(2006)\citenamefont
  {Trompeter}, \citenamefont {Pertsch}, \citenamefont {Lederer}, \citenamefont
  {Michaelis}, \citenamefont {Streppel}, \citenamefont {Br\"auer},\ and\
  \citenamefont {Peschel}}]{trompeter2006visual}%
  \BibitemOpen
  \bibfield  {author} {\bibinfo {author} {\bibfnamefont {H.}~\bibnamefont
  {Trompeter}}, \bibinfo {author} {\bibfnamefont {T.}~\bibnamefont {Pertsch}},
  \bibinfo {author} {\bibfnamefont {F.}~\bibnamefont {Lederer}}, \bibinfo
  {author} {\bibfnamefont {D.}~\bibnamefont {Michaelis}}, \bibinfo {author}
  {\bibfnamefont {U.}~\bibnamefont {Streppel}}, \bibinfo {author}
  {\bibfnamefont {A.}~\bibnamefont {Br\"auer}}, \ and\ \bibinfo {author}
  {\bibfnamefont {U.}~\bibnamefont {Peschel}},\ }\href {\doibase
  10.1103/PhysRevLett.96.023901} {\bibfield  {journal} {\bibinfo  {journal}
  {Phys. Rev. Lett.}\ }\textbf {\bibinfo {volume} {96}},\ \bibinfo {pages}
  {023901} (\bibinfo {year} {2006})}\BibitemShut {NoStop}%
\bibitem [{\citenamefont {Perez-Leija}\ \emph {et~al.}(2016)\citenamefont
  {Perez-Leija}, \citenamefont {Szameit}, \citenamefont {Ramos-Prieto},
  \citenamefont {Moya-Cessa},\ and\ \citenamefont
  {Christodoulides}}]{perez2016generalized}%
  \BibitemOpen
  \bibfield  {author} {\bibinfo {author} {\bibfnamefont {A.}~\bibnamefont
  {Perez-Leija}}, \bibinfo {author} {\bibfnamefont {A.}~\bibnamefont
  {Szameit}}, \bibinfo {author} {\bibfnamefont {I.}~\bibnamefont
  {Ramos-Prieto}}, \bibinfo {author} {\bibfnamefont {H.}~\bibnamefont
  {Moya-Cessa}}, \ and\ \bibinfo {author} {\bibfnamefont {D.~N.}\ \bibnamefont
  {Christodoulides}},\ }\href {\doibase 10.1103/PhysRevA.93.053815} {\bibfield
  {journal} {\bibinfo  {journal} {Phys. Rev. A}\ }\textbf {\bibinfo {volume}
  {93}},\ \bibinfo {pages} {053815} (\bibinfo {year} {2016})}\BibitemShut
  {NoStop}%
\bibitem [{\citenamefont {Rom\'an-Ancheyta}\ \emph {et~al.}(2017)\citenamefont
  {Rom\'an-Ancheyta}, \citenamefont {Ramos-Prieto}, \citenamefont
  {Perez-Leija}, \citenamefont {Busch},\ and\ \citenamefont
  {Le\'on-Montiel}}]{Ancheyta_2017}%
  \BibitemOpen
  \bibfield  {author} {\bibinfo {author} {\bibfnamefont {R.}~\bibnamefont
  {Rom\'an-Ancheyta}}, \bibinfo {author} {\bibfnamefont {I.}~\bibnamefont
  {Ramos-Prieto}}, \bibinfo {author} {\bibfnamefont {A.}~\bibnamefont
  {Perez-Leija}}, \bibinfo {author} {\bibfnamefont {K.}~\bibnamefont {Busch}},
  \ and\ \bibinfo {author} {\bibfnamefont {R.~d.~J.}\ \bibnamefont
  {Le\'on-Montiel}},\ }\href {\doibase 10.1103/PhysRevA.96.032501} {\bibfield
  {journal} {\bibinfo  {journal} {Phys. Rev. A}\ }\textbf {\bibinfo {volume}
  {96}},\ \bibinfo {pages} {032501} (\bibinfo {year} {2017})}\BibitemShut
  {NoStop}%
\bibitem [{\citenamefont {Ramos-Prieto}\ \emph {et~al.}(2021)\citenamefont
  {Ramos-Prieto}, \citenamefont {Uriostegui}, \citenamefont {R\'{e}camier},
  \citenamefont {Soto-Eguibar},\ and\ \citenamefont {Moya-Cessa}}]{Ramos_2021}%
  \BibitemOpen
  \bibfield  {author} {\bibinfo {author} {\bibfnamefont {I.}~\bibnamefont
  {Ramos-Prieto}}, \bibinfo {author} {\bibfnamefont {K.}~\bibnamefont
  {Uriostegui}}, \bibinfo {author} {\bibfnamefont {J.}~\bibnamefont
  {R\'{e}camier}}, \bibinfo {author} {\bibfnamefont {F.}~\bibnamefont
  {Soto-Eguibar}}, \ and\ \bibinfo {author} {\bibfnamefont {H.~M.}\
  \bibnamefont {Moya-Cessa}},\ }\href {\doibase 10.1364/OL.437829} {\bibfield
  {journal} {\bibinfo  {journal} {Opt. Lett.}\ }\textbf {\bibinfo {volume}
  {46}},\ \bibinfo {pages} {4690} (\bibinfo {year} {2021})}\BibitemShut
  {NoStop}%
\bibitem [{\citenamefont {Urzúa}\ \emph {et~al.}(2024)\citenamefont {Urzúa},
  \citenamefont {Ramos-Prieto},\ and\ \citenamefont {Moya-Cessa}}]{Urzua_2024}%
  \BibitemOpen
  \bibfield  {author} {\bibinfo {author} {\bibfnamefont {A.~R.}\ \bibnamefont
  {Urzúa}}, \bibinfo {author} {\bibfnamefont {I.}~\bibnamefont
  {Ramos-Prieto}}, \ and\ \bibinfo {author} {\bibfnamefont {H.~M.}\
  \bibnamefont {Moya-Cessa}},\ }\href {https://arxiv.org/abs/2406.15641}
  {\enquote {\bibinfo {title} {Integrated optical wave analyzer using the
  discrete fractional fourier transform},}\ } (\bibinfo {year} {2024}),\
  \Eprint {http://arxiv.org/abs/2406.15641} {arXiv:2406.15641 [physics.optics]}
  \BibitemShut {NoStop}%
\bibitem [{\citenamefont {Altland}\ and\ \citenamefont
  {Simons}(2010)}]{altland2010condensed}%
  \BibitemOpen
  \bibfield  {author} {\bibinfo {author} {\bibfnamefont {A.}~\bibnamefont
  {Altland}}\ and\ \bibinfo {author} {\bibfnamefont {B.}~\bibnamefont
  {Simons}},\ }\href {https://books.google.com.mx/books?id=gk4hAwAAQBAJ} {\emph
  {\bibinfo {title} {Condensed Matter Field Theory}}}\ (\bibinfo  {publisher}
  {Cambridge University Press},\ \bibinfo {year} {2010})\BibitemShut {NoStop}%
\bibitem [{\citenamefont {Kartashov}\ \emph {et~al.}(2011)\citenamefont
  {Kartashov}, \citenamefont {Malomed},\ and\ \citenamefont
  {Torner}}]{kartashov2011solitons}%
  \BibitemOpen
  \bibfield  {author} {\bibinfo {author} {\bibfnamefont {Y.~V.}\ \bibnamefont
  {Kartashov}}, \bibinfo {author} {\bibfnamefont {B.~A.}\ \bibnamefont
  {Malomed}}, \ and\ \bibinfo {author} {\bibfnamefont {L.}~\bibnamefont
  {Torner}},\ }\href {\doibase 10.1103/RevModPhys.83.247} {\bibfield  {journal}
  {\bibinfo  {journal} {Rev. Mod. Phys.}\ }\textbf {\bibinfo {volume} {83}},\
  \bibinfo {pages} {247} (\bibinfo {year} {2011})}\BibitemShut {NoStop}%
\bibitem [{\citenamefont {Sch{\"a}fer}\ \emph {et~al.}(2020)\citenamefont
  {Sch{\"a}fer}, \citenamefont {Fukuhara}, \citenamefont {Sugawa},
  \citenamefont {Takasu},\ and\ \citenamefont {Takahashi}}]{schafer2020tools}%
  \BibitemOpen
  \bibfield  {author} {\bibinfo {author} {\bibfnamefont {F.}~\bibnamefont
  {Sch{\"a}fer}}, \bibinfo {author} {\bibfnamefont {T.}~\bibnamefont
  {Fukuhara}}, \bibinfo {author} {\bibfnamefont {S.}~\bibnamefont {Sugawa}},
  \bibinfo {author} {\bibfnamefont {Y.}~\bibnamefont {Takasu}}, \ and\ \bibinfo
  {author} {\bibfnamefont {Y.}~\bibnamefont {Takahashi}},\ }\href {\doibase
  10.1038/s42254-020-0195-3} {\bibfield  {journal} {\bibinfo  {journal} {Nat.
  Rev. Phys.}\ }\textbf {\bibinfo {volume} {2}},\ \bibinfo {pages} {411}
  (\bibinfo {year} {2020})}\BibitemShut {NoStop}%
\bibitem [{\citenamefont {Lewenstein}\ \emph {et~al.}(2012)\citenamefont
  {Lewenstein}, \citenamefont {Sanpera},\ and\ \citenamefont
  {Ahufinger}}]{lewenstein2012ultracold}%
  \BibitemOpen
  \bibfield  {author} {\bibinfo {author} {\bibfnamefont {M.}~\bibnamefont
  {Lewenstein}}, \bibinfo {author} {\bibfnamefont {A.}~\bibnamefont {Sanpera}},
  \ and\ \bibinfo {author} {\bibfnamefont {V.}~\bibnamefont {Ahufinger}},\
  }\href {https://books.google.com.mx/books?id=Wpl91RDxV5IC} {\emph {\bibinfo
  {title} {Ultracold Atoms in Optical Lattices: Simulating quantum many-body
  systems}}}\ (\bibinfo  {publisher} {OUP Oxford},\ \bibinfo {year}
  {2012})\BibitemShut {NoStop}%
\bibitem [{\citenamefont {Man'ko}\ \emph {et~al.}(1997)\citenamefont {Man'ko},
  \citenamefont {Marmo}, \citenamefont {Sudarshan},\ and\ \citenamefont
  {Zaccaria}}]{man1997f}%
  \BibitemOpen
  \bibfield  {author} {\bibinfo {author} {\bibfnamefont {V.~I.}\ \bibnamefont
  {Man'ko}}, \bibinfo {author} {\bibfnamefont {G.}~\bibnamefont {Marmo}},
  \bibinfo {author} {\bibfnamefont {E.~C.~G.}\ \bibnamefont {Sudarshan}}, \
  and\ \bibinfo {author} {\bibfnamefont {F.}~\bibnamefont {Zaccaria}},\ }\href
  {\doibase 10.1088/0031-8949/55/5/004} {\bibfield  {journal} {\bibinfo
  {journal} {Phys. Scr.}\ }\textbf {\bibinfo {volume} {55}},\ \bibinfo {pages}
  {528} (\bibinfo {year} {1997})}\BibitemShut {NoStop}%
\bibitem [{\citenamefont {de~los Santos-Sánchez}\ \emph
  {et~al.}(2015)\citenamefont {de~los Santos-Sánchez}, \citenamefont
  {Récamier},\ and\ \citenamefont {Jáuregui}}]{de2015markovian}%
  \BibitemOpen
  \bibfield  {author} {\bibinfo {author} {\bibfnamefont {O.}~\bibnamefont
  {de~los Santos-Sánchez}}, \bibinfo {author} {\bibfnamefont {J.}~\bibnamefont
  {Récamier}}, \ and\ \bibinfo {author} {\bibfnamefont {R.}~\bibnamefont
  {Jáuregui}},\ }\href {\doibase 10.1088/0031-8949/90/7/074018} {\bibfield
  {journal} {\bibinfo  {journal} {Phys. Scr.}\ }\textbf {\bibinfo {volume}
  {90}},\ \bibinfo {pages} {074018} (\bibinfo {year} {2015})}\BibitemShut
  {NoStop}%
\bibitem [{\citenamefont {de~los Santos-Sánchez}\ and\ \citenamefont
  {Récamier}(2011)}]{de2011f}%
  \BibitemOpen
  \bibfield  {author} {\bibinfo {author} {\bibfnamefont {O.}~\bibnamefont
  {de~los Santos-Sánchez}}\ and\ \bibinfo {author} {\bibfnamefont
  {J.}~\bibnamefont {Récamier}},\ }\href {\doibase
  10.1088/0953-4075/45/1/015502} {\bibfield  {journal} {\bibinfo  {journal} {J.
  Phys. B: At. Mol. Opt. Phys}\ }\textbf {\bibinfo {volume} {45}},\ \bibinfo
  {pages} {015502} (\bibinfo {year} {2011})}\BibitemShut {NoStop}%
\bibitem [{\citenamefont {Ramos-Prieto}\ \emph {et~al.}(2014)\citenamefont
  {Ramos-Prieto}, \citenamefont {Rodr\'{\i}guez-Lara},\ and\ \citenamefont
  {Moya-Cessa}}]{Ramos_2014}%
  \BibitemOpen
  \bibfield  {author} {\bibinfo {author} {\bibfnamefont {I.}~\bibnamefont
  {Ramos-Prieto}}, \bibinfo {author} {\bibfnamefont {B.~M.}\ \bibnamefont
  {Rodr\'{\i}guez-Lara}}, \ and\ \bibinfo {author} {\bibfnamefont {H.~M.}\
  \bibnamefont {Moya-Cessa}},\ }\href {\doibase 10.1142/S0219749915600059}
  {\bibfield  {journal} {\bibinfo  {journal} {International Journal of Quantum
  Information}\ }\textbf {\bibinfo {volume} {12}},\ \bibinfo {pages} {1560005}
  (\bibinfo {year} {2014})}\BibitemShut {NoStop}%
\bibitem [{\citenamefont {London}(1926)}]{london1926jacobischen}%
  \BibitemOpen
  \bibfield  {author} {\bibinfo {author} {\bibfnamefont {F.}~\bibnamefont
  {London}},\ }\href {\doibase 10.1007/BF01397484} {\bibfield  {journal}
  {\bibinfo  {journal} {Z. Physik}\ }\textbf {\bibinfo {volume} {37}},\
  \bibinfo {pages} {915} (\bibinfo {year} {1926})}\BibitemShut {NoStop}%
\bibitem [{\citenamefont {Susskind}\ and\ \citenamefont
  {Glogower}(1964)}]{susskind1964quantum}%
  \BibitemOpen
  \bibfield  {author} {\bibinfo {author} {\bibfnamefont {L.}~\bibnamefont
  {Susskind}}\ and\ \bibinfo {author} {\bibfnamefont {J.}~\bibnamefont
  {Glogower}},\ }\href {\doibase 10.1103/PhysicsPhysiqueFizika.1.49} {\bibfield
   {journal} {\bibinfo  {journal} {Physics Physique Fizika}\ }\textbf {\bibinfo
  {volume} {1}},\ \bibinfo {pages} {49} (\bibinfo {year} {1964})}\BibitemShut
  {NoStop}%
\bibitem [{\citenamefont {Abramowitz}\ and\ \citenamefont
  {Stegun}(1968)}]{abramowitz1968handbook}%
  \BibitemOpen
  \bibfield  {author} {\bibinfo {author} {\bibfnamefont {M.}~\bibnamefont
  {Abramowitz}}\ and\ \bibinfo {author} {\bibfnamefont {I.}~\bibnamefont
  {Stegun}},\ }\href {https://books.google.com.mx/books?id=ZboM5tOFWtsC} {\emph
  {\bibinfo {title} {Handbook of Mathematical Functions with Formulas, Graphs,
  and Mathematical Tables}}},\ \bibinfo {series} {Applied mathematics series}\
  No.\ \bibinfo {number} {v. 55, no. 1972}\ (\bibinfo  {publisher} {U.S.
  Government Printing Office},\ \bibinfo {year} {1968})\BibitemShut {NoStop}%
\bibitem [{\citenamefont {Arfken}\ \emph {et~al.}(2011)\citenamefont {Arfken},
  \citenamefont {Weber},\ and\ \citenamefont
  {Harris}}]{arfken2011mathematical}%
  \BibitemOpen
  \bibfield  {author} {\bibinfo {author} {\bibfnamefont {G.}~\bibnamefont
  {Arfken}}, \bibinfo {author} {\bibfnamefont {H.}~\bibnamefont {Weber}}, \
  and\ \bibinfo {author} {\bibfnamefont {F.}~\bibnamefont {Harris}},\ }\href
  {https://books.google.com.mx/books?id=JOpHkJF-qcwC} {\emph {\bibinfo {title}
  {Mathematical Methods for Physicists: A Comprehensive Guide}}}\ (\bibinfo
  {publisher} {Elsevier Science},\ \bibinfo {year} {2011})\BibitemShut
  {NoStop}%
\bibitem [{\citenamefont {Gerry}\ and\ \citenamefont
  {Knight}(2023)}]{gerry2023introductory}%
  \BibitemOpen
  \bibfield  {author} {\bibinfo {author} {\bibfnamefont {C.}~\bibnamefont
  {Gerry}}\ and\ \bibinfo {author} {\bibfnamefont {P.}~\bibnamefont {Knight}},\
  }\href {https://books.google.com.mx/books?id=p-_kEAAAQBAJ} {\emph {\bibinfo
  {title} {Introductory Quantum Optics}}}\ (\bibinfo  {publisher} {Cambridge
  University Press},\ \bibinfo {year} {2023})\BibitemShut {NoStop}%
\bibitem [{\citenamefont {Gerry}\ and\ \citenamefont
  {Mimih}(2010)}]{gerry2010parity}%
  \BibitemOpen
  \bibfield  {author} {\bibinfo {author} {\bibfnamefont {C.~C.}\ \bibnamefont
  {Gerry}}\ and\ \bibinfo {author} {\bibfnamefont {J.}~\bibnamefont {Mimih}},\
  }\href {\doibase 10.1080/00107514.2010.509995} {\bibfield  {journal}
  {\bibinfo  {journal} {Contemporary Physics}\ }\textbf {\bibinfo {volume}
  {51}},\ \bibinfo {pages} {497} (\bibinfo {year} {2010})}\BibitemShut
  {NoStop}%
\bibitem [{\citenamefont {Sakurai}\ and\ \citenamefont
  {Napolitano}(2018)}]{sakurai2020modern}%
  \BibitemOpen
  \bibfield  {author} {\bibinfo {author} {\bibfnamefont {J.~J.}\ \bibnamefont
  {Sakurai}}\ and\ \bibinfo {author} {\bibfnamefont {J.}~\bibnamefont
  {Napolitano}},\ }\href {\doibase https://doi.org/10.1017/9781108499996}
  {\emph {\bibinfo {title} {Modern quantum mechanics}}}\ (\bibinfo  {publisher}
  {Cambridge University Press},\ \bibinfo {year} {2018})\BibitemShut {NoStop}%
\bibitem [{\citenamefont {Jordana}\ \emph {et~al.}(2007)\citenamefont
  {Jordana}, \citenamefont {Fedeli}, \citenamefont {Lyan}, \citenamefont
  {Colonna}, \citenamefont {Gautier}, \citenamefont {Daldosso}, \citenamefont
  {Pavesi}, \citenamefont {Lebour}, \citenamefont {Pellegrino}, \citenamefont
  {Garrido}, \citenamefont {Blasco}, \citenamefont {Cuesta-Soto},\ and\
  \citenamefont {Sanchis}}]{jordana2007deep}%
  \BibitemOpen
  \bibfield  {author} {\bibinfo {author} {\bibfnamefont {E.}~\bibnamefont
  {Jordana}}, \bibinfo {author} {\bibfnamefont {J.-M.}\ \bibnamefont {Fedeli}},
  \bibinfo {author} {\bibfnamefont {P.}~\bibnamefont {Lyan}}, \bibinfo {author}
  {\bibfnamefont {J.}~\bibnamefont {Colonna}}, \bibinfo {author} {\bibfnamefont
  {P.}~\bibnamefont {Gautier}}, \bibinfo {author} {\bibfnamefont
  {N.}~\bibnamefont {Daldosso}}, \bibinfo {author} {\bibfnamefont
  {L.}~\bibnamefont {Pavesi}}, \bibinfo {author} {\bibfnamefont
  {Y.}~\bibnamefont {Lebour}}, \bibinfo {author} {\bibfnamefont
  {P.}~\bibnamefont {Pellegrino}}, \bibinfo {author} {\bibfnamefont
  {B.}~\bibnamefont {Garrido}}, \bibinfo {author} {\bibfnamefont
  {J.}~\bibnamefont {Blasco}}, \bibinfo {author} {\bibfnamefont
  {F.}~\bibnamefont {Cuesta-Soto}}, \ and\ \bibinfo {author} {\bibfnamefont
  {P.}~\bibnamefont {Sanchis}},\ }in\ \href {\doibase
  10.1109/GROUP4.2007.4347722} {\emph {\bibinfo {booktitle} {2007 4th IEEE
  International Conference on Group IV Photonics}}}\ (\bibinfo {year} {2007})\
  pp.\ \bibinfo {pages} {1--3}\BibitemShut {NoStop}%
\bibitem [{\citenamefont {Szameit}\ \emph {et~al.}(2007)\citenamefont
  {Szameit}, \citenamefont {Dreisow}, \citenamefont {Pertsch}, \citenamefont
  {Nolte},\ and\ \citenamefont {T\"{u}nnermann}}]{szameit2007control}%
  \BibitemOpen
  \bibfield  {author} {\bibinfo {author} {\bibfnamefont {A.}~\bibnamefont
  {Szameit}}, \bibinfo {author} {\bibfnamefont {F.}~\bibnamefont {Dreisow}},
  \bibinfo {author} {\bibfnamefont {T.}~\bibnamefont {Pertsch}}, \bibinfo
  {author} {\bibfnamefont {S.}~\bibnamefont {Nolte}}, \ and\ \bibinfo {author}
  {\bibfnamefont {A.}~\bibnamefont {T\"{u}nnermann}},\ }\href {\doibase
  10.1364/OE.15.001579} {\bibfield  {journal} {\bibinfo  {journal} {Opt.
  Express}\ }\textbf {\bibinfo {volume} {15}},\ \bibinfo {pages} {1579}
  (\bibinfo {year} {2007})}\BibitemShut {NoStop}%
\bibitem [{\citenamefont {Bojko}\ \emph {et~al.}(2011)\citenamefont {Bojko},
  \citenamefont {Li}, \citenamefont {He}, \citenamefont {Baehr-Jones},
  \citenamefont {Hochberg},\ and\ \citenamefont {Aida}}]{bojko2011electron}%
  \BibitemOpen
  \bibfield  {author} {\bibinfo {author} {\bibfnamefont {R.~J.}\ \bibnamefont
  {Bojko}}, \bibinfo {author} {\bibfnamefont {J.}~\bibnamefont {Li}}, \bibinfo
  {author} {\bibfnamefont {L.}~\bibnamefont {He}}, \bibinfo {author}
  {\bibfnamefont {T.}~\bibnamefont {Baehr-Jones}}, \bibinfo {author}
  {\bibfnamefont {M.}~\bibnamefont {Hochberg}}, \ and\ \bibinfo {author}
  {\bibfnamefont {Y.}~\bibnamefont {Aida}},\ }\href {\doibase
  10.1116/1.3653266} {\bibfield  {journal} {\bibinfo  {journal} {Journal of
  Vacuum Science \& Technology B}\ }\textbf {\bibinfo {volume} {29}},\ \bibinfo
  {pages} {06F309} (\bibinfo {year} {2011})}\BibitemShut {NoStop}%
\end{thebibliography}

%

\end{document}